\documentclass[12pt]{article}
\usepackage{bm,bbm,lscape}
\usepackage{natbib, amsmath,authblk}
\usepackage{marvosym}
\usepackage{multicol}
\usepackage[british]{babel}
\usepackage{pstricks,graphicx,xcolor,geometry,setspace,lineno}
\usepackage[british]{babel}
\usepackage[justification=justified,singlelinecheck=false,labelfont=it]{caption}
\usepackage{pstricks,graphicx,eurosym}

\title{Kernel-based semiparametric multinomial logit modelling of political party affiliation}
 
 \author[1]{Roland Langrock\footnote{Corresponding author. Email: roland@mcs.st-and.ac.uk. Phone: +44 (0) 1334 46 1829.}}
 \author[2]{Nils-Bastian Heidenreich}
 \author[3]{Stefan Sperlich}

 \affil[1]{\small University of St Andrews, UK}
 \affil[2]{\small Georg-August Universit{\"a}t G{\"o}ttingen, Germany}
 \affil[3]{\small Universit\'e de Gen\`eve, Switzerland}

\date{}

\begin{document}

\maketitle

\begin{abstract}
Conventional, parametric multinomial logit models are in general not sufficient for detecting the complex patterns voter profiles nowadays typically exhibit. In this manuscript, we use a semiparametric multinomial logit model to give a detailed analysis of the composition of a subsample of the German electorate in 2006. Germany is a particularly strong case for more flexible nonparametric approaches in this context, since due to the reunification and the preceding different political histories the composition of the electorate is very complex and nuanced. Our analysis reveals strong interactions of the covariates age and income, and highly nonlinear shapes of the factor impacts for each party's likelihood to be voted. Notably, we develop and provide a smoothed likelihood estimator for semiparametric multinomial logit models, which can be applied also in other application fields, such as, e.g., marketing. \\[1em]
\bigskip
\end{abstract}

\begin{spacing}{1}

\section{Introduction and Motivation}

The multinomial logit (MNL) model
allows to investigate the influence of a vector of covariates on more than two possibly unordered
outcomes of categorical response variables. It became a popular tool in econometrics following the work on brand
choice behaviour by McFadden (1974) and on urban travel demand by Domencich and McFadden
(1975), respectively. MNL regression also provides a means of analysing how socio-economic factors and other covariates 
affect an individual's likelihood of supporting various political parties. 
Such information is of great importance for policy makers and analysts, for example when it comes 
to designing of campaigns for targeted voter groups.  
Regression models can also be useful in forecasting election outcomes from opinion polls, 
and in particular from exit polls (Curtice and Firth 2008; Fisher et al.\ 2011).
However, we claim that conventional, parametric MNL models are in general not 
adequate for capturing the complex patterns typically exhibited by voter profiles. 

In this manuscript, we use data on political party affiliation in Germany from 2006 
to demonstrate the usefulness of a semiparametric extension of the basic MNL model in this context.
We compare our results to corresponding results obtained from 
fitting a parametric MNL model, thereby demonstrating the superiority of the more flexible semiparametric 
approach when analyzing this kind of data. The increase in flexibility offered by the semiparametric approach 
allows us to identify complex voter profiles with respect to age, 
income, gender and region, with high nonlinearity in the covariate effects and strong interaction 
between different covariates. 
This data structure can most naturally be modelled by considering multidimensional nonparametric functions of covariates. 
We explore a model fitting approach based on kernel smoothing and a profile likelihood algorithm. The key accomplishments of this manuscript thus are:
1) a detailed analysis of a complex electorate, demonstrating the usefulness of flexible regression models in this context, and 2) 
the presentation of an original, kernel-based approach for the estimation of semiparametric MNL models. 

Different trials were previously undertaken to incorporate nonlinear effects of the explanatory variables into multiple choice models. 
Krishnamurthi and Raj (1988) used logarithmic transformations. 
Ben-Akiva and Lerman (1985) as well as Kalyanaram and Little
(1994) proposed using piecewise linear (utility) functions on predetermined (sub)intervals.
More recently, some authors developed nonparametric and semiparametric methods for 
regression models with multicategorical response. Yee and Wild (1996) considered a backfitting algorithm on a class of
multivariate additive models using smoothing splines. Abe (1998, 1999) proposed a special
class of generalized additive models which accommodates a multinomial qualitative
response. His algorithm is based on a penalised likelihood
function and modified local scoring (cf.\ Hastie and Tibshirani 1986). Tutz and Scholz (2004)
approximate unspecified additive functions by a finite number of basis functions which
are penalised with respect to their localisation. Kneib et al.\ (2007)
modified this using penalised B-splines and a Bayesian approach for their estimation.
In this manuscript, we explore an alternative approach using a likelihood that is localised via
kernels. More specifically, we consider profile likelihood estimation in the spirit of
Severini and Wong (1992). Our methods extend the work by M{\"u}ller (2001) -- 
who considered, {\em inter alia}, the estimation of semiparametric models with
binary response variables -- to the case of multinomial responses.
 
The paper is structured as follows. The motivating data set on political party affiliation is introduced in Section \ref{sec-appl1}. 
In Section \ref{sec-model}, we formulate the semiparametric MNL model and describe the kernel-based
estimation procedure. The results of fitting both fully parametric and semiparametric MNL models to the data on political party affiliation are given in Section \ref{results}. The parametric MNL analysis here serves as a benchmark for the results of the semiparametric approach. Concluding remarks are given in Section \ref{sec-concl}.

\section{The data on political party affiliation in Germany} \label{sec-appl1}

We first introduce the motivating data set on political party affiliation in Germany. 
Our aim is to identify typical voter groups of the dominant 
political parties in the multi-party system of Germany at the time of the data collection, i.e., in 2006.
In order to make subsequent interpretations accessible to a broad readership, 
we begin by briefly sketching the current composition of the German party system and the historical background.

Nowadays, the German party system comprises five main political parties: the
Christian Democratic Union and its sister party, the Bavarian Christian
Social Union, which together form one liberal-conservative parliamentary group (CU henceforth), the Social
Democratic Party (SPD), the liberal Free Democratic Party (FDP), the democratic socialist Left Party (LP) and the green party Alliance~'90/The
Greens (A90G), and it is these five parties that we will focus on in subsequent considerations.

From the establishment of the German Bundestag (i.e., the German parliament) in 1949 until the German reunification in 1990, 
Western Germany was governed either by the CU or the SPD,
with absolute majority or in a coalition with the FDP. In the seventies, diverse groups of
alternative green activists contested at various local elections, and in 1980 a green
organisation was founded at the federal level. It comprised groups such as
the anti-nuclear movement, the student movement, feminist
groups and the peace movement (L\"osche 1993). They won the first seats in
the German Bundestag in 1983, and from 1998 to 2005 they formed the government in
a coalition with the SPD. In the East German dictatorship prior to the reunification, the Socialist Unity Party (SED) had sole political power,
although small and well-controlled Christian and liberal parties co-existed to give the
system a semblance of legitimacy. After reunification the so-called PDS was founded as
the heir of the SED. In 2005, the PDS entered an alliance with the then founded Western German
party ``Labour and Social Justice - Electoral Alternative (WASG)''. Since 2007 the alliance is simply called
``The Left'' (LP throughout this manuscript). 

The data for the subsequent analysis was extracted from the German Socio-Economic Panel (SOEP) of the year 2006. 
The variable of interest is political party affiliation, i.e., the answer to the question ``Towards which party do you
lean?''. As is commonly done in the literature, the socio-economic factors that we
consider are age, log-income (i.e., the logarithm of the monthly net total household income),
region of origin, and gender (Dow and Endersby 2004; Quinn et al.\ 1999).
The variable associated with the region of origin is a dummy 
indicating whether or not a person was resided in Eastern Germany {\em before} reunification.
The principal aim of this manuscript is to illustrate the usefulness of the proposed modelling approach in voter profiling in general. Therefore, 
we abstain from considering additional covariates, such as, e.g., religion or education, and from carrying out a model selection exercise to identify the most relevant covariates.

In total, 8787 individuals reported their party affiliation in the original data set.
From this data set we excluded 376 (4.3\%) individuals who favoured a party different from the five main parties that we focus on, 
and 426 (4.8\%) individuals who made an implausible or no declaration about their income. 
In the remaining subsample, 227 persons who lived abroad before reunification, or did not report 
their region of origin, were assigned to Western Germany
(the results remained virtually identical when we excluded them). 
The descriptive statistics for the considered sample are
summarized in Tables \ref{Desc2} and \ref{Desc1}. Notably, the income distribution is strongly skewed to
the right.   

\begin{table}[htb]
\caption[Table]{Descriptive statistics for the considered covariates: gender (1 if female), region of origin (1 if from Eastern Germany), income (monthly net total household income in Euro) and age (in years).  \\ {\white 0}}
\label{Desc2}
\tabcolsep4pt
\begin{center}
\begin{small}
\begin{tabular}{lccccc}
Variable & & \\
\hline\\[-0.9em]
       & no.\ of 0s & no.\ of 1s  \\
gender & 4\,063 & 3\,922  \\
region & 6\,267 & 1\,718   \\[0.5em]
    & min  & max  & mean & median &\\
income   &  400 & 30,000  & 3,089  & 2,600 &\\
age   &  21 & 97 & 53.8  & 54.0 &
\end{tabular}
\end{small}
\end{center}
\end{table}

\begin{table}[htb]
\caption[Table]{Percentages of reported political affiliation. \\ {\white 0}}
\label{Desc1}
\tabcolsep4pt
\begin{center}
\begin{small}
\begin{tabular}{llccccc}
Party & & CU & SPD & A90G & LP & FDP \\
\hline\\[-0.9em]
Affiliation & (in \%)& 42.17 & 37.86 & 8.87 & 6.35 & 4.74
\end{tabular}
\end{small}
\end{center}
\end{table}
 
There are two substantially different ways to analyse voter groups: 1) using purely
descriptive statistics based on public-opinion polls, as routinely published by market
research institutes such as Infratest dimap or Forsa in Germany, and 2) employing
inferential statistics by fitting adequate models. When it comes to voter profiling, one
of the main drawbacks of 1) is that the distribution of the voter's choice is not
quantified based on statistical laws and hence can not be used to support inferential
statements about the population. Furthermore, such analyses typically focus on only one or
two covariates at a time. Models such as the MNL model
attempt to overcome these deficiencies by modelling the voter's party affiliation
as outcome of a distribution that depends on a number of covariates.
However, the MNL model and similar parametric models also have limitations:
they are based on assumptions concerning the specific functional form that links the
covariates  to the outcome. Another limitation is the commonly assumed additive separability and the 
implied neglect of possible interactions between different covariates. 
These aspects will be studied in detail in Section \ref{results}, 
where we give the results of fitting MNL models to the data described here. But in the next section, 
we first of all introduce the semiparametric MNL model 
and describe its estimation based on kernels and profile likelihood.

\section{The semiparametric MNL model and its kernel-based estimation} \label{sec-model}

\subsection{Model formulation} \label{sec-modela}

We consider a semiparametric MNL model with $K$ different outcome categories
that have no natural order. The conditional probability of outcome $Y=k$, $k=1,\ldots,K$,
given the individual covariate vectors ${\mathbf X}=(X_1, \ldots, X_p)^t\in\mathbbm{R}^p$ and
${\mathbf T}=(T_1, \ldots, T_q)^t\in\mathbbm{R}^q$, is assumed to be given by
\begin{equation}\label{model}
\mathbbm{P}(Y=k \, \vert \, {\mathbf X},{\mathbf T})=\frac{\exp
\left( {\mathbf X}^t \boldsymbol{\beta}_k + m_k({\mathbf T})
\right)}{\sum_{j=1}^K \exp \left( {\mathbf X}^t
\boldsymbol{\beta}_j + m_j({\mathbf T}) \right) } \, .
\end{equation}
To ensure identifiability we set $\boldsymbol{\beta}_K={\mathbf 0}$, and $m_K =
0$, such that\ $K$ is the reference category.  Each $m_k(\cdot)$, $k=1,\ldots,K-1$, is assumed to be a smooth
function with domain $\mathbbm{R}^q$ and each
$\boldsymbol{\beta}_k=(\beta_{k1},\ldots,\beta_{kp})^t$, $k=1,\ldots,K-1$, denotes an
unknown parameter vector.

\subsection{Estimation via kernels and profile likelihood} \label{sec-modelb}

If the functions $m_k(\cdot)$ were known it would be easy to find estimators for the
vectors $\boldsymbol{\beta}_k$, and vice versa. Following the concept of profile
likelihood (1992), the functions $m_k(\cdot)$ are regarded as nuisance
when estimating the finite-dimensional parameters $\boldsymbol{\beta}_k$. The functions $m_k(\cdot)$ themselves
can be estimated via kernel smoothing. Note that the estimate of
$m_k(\cdot)$ will depend on all
$\boldsymbol{\beta}_j$, $j=1,\ldots ,K-1$, which in the following is indicated using the notation $m_{k,\boldsymbol{\beta}_{{\displaystyle \cdot}}}(\cdot)$. 
Under regularity conditions, one obtains asymptotically normal,
$\sqrt{n}$-consistent and efficient estimators for the vectors $\boldsymbol{\beta}_k$
owing to likelihood estimation.   
For the functions $m_k$, $k=1,\ldots,K-1$, one obtains consistent estimators with
statistical properties typical for nonparametric kernel smoothing.

In order to estimate the so-called least favourable curve
$m_{k,\boldsymbol{\beta}_{{\displaystyle \cdot}}}(\mathbf
t)$ at point ${\mathbf
t}:=(t_1,\ldots,t_q)$ for given $\boldsymbol{\beta}_j$, $j=1,\ldots,K-1$, we use a
$q$-dimensional kernel $\mathcal{K}: \mathbbm{R}^q \rightarrow \mathbbm{R}$, a bandwidth matrix
${\mathbf H}\in \mathbbm{R}_+^{q\times q}$, and consider the local likelihood
\begin{eqnarray}\label{smolik}
{\mathcal{L}}_s (m_{k,\boldsymbol{\beta}_{{\displaystyle \cdot}}}({\mathbf t}))
= \sum_{i=1}^n (\det \mathbf H)^{-1} \mathcal{K} \bigl( {\mathbf H}^{-1} ({\mathbf t}-{\mathbf
t}_i) \bigr)  {\mathcal{L}}
(\boldsymbol{\eta}_{i}(m_{k,\boldsymbol{\beta}_{{\displaystyle \cdot}}}({\mathbf
t})),y_i), & & 
\\ \nonumber    \mbox{with \, }
\boldsymbol{\eta}_{i}(m_{k,\boldsymbol{\beta}_{{\displaystyle \cdot}}}({\mathbf t}))
:=(\eta_{1i},\ldots,\eta_{ki}(m_{k,\boldsymbol{\beta}_{{\displaystyle \cdot}}}({\mathbf t}))
,\ldots,\eta_{Ki}), & &  \\ \nonumber   \mbox{where \, }
\eta_{ki}(m_{k,\boldsymbol{\beta}_{{\displaystyle \cdot}}}({\mathbf t}))
:={\mathbf x}_i^t \boldsymbol{\beta}_k +
m_{k,\boldsymbol{\beta}_{{\displaystyle \cdot}}}({\mathbf t}) & &   \\ \nonumber \mbox{and \, }
\eta_{ji} :={\mathbf x}_i^t \boldsymbol{\beta}_j +
m_{j,\boldsymbol{\beta}_{{\displaystyle \cdot}}}({\mathbf t}_i) \quad \text{for} \; j
\neq k, & & 
\end{eqnarray}
with row vector ${\mathbf x}_i^t=(x_{i1},\ldots,x_{ip})$ and ${\mathbf t}_i=(t_{i1},\ldots,t_{iq})$.
Here ${\mathcal{L}} (\boldsymbol{\eta}_{i}(m_{k,\boldsymbol{\beta}_{{\displaystyle
\cdot}}}({\mathbf t})),y_i)$ denotes the log-likelihood of (\ref{model}) of the $i$th
observation with predictor $\boldsymbol{\eta}_{i}(m_{k,\boldsymbol{\beta}_{{\displaystyle
\cdot}}}({\mathbf t}))$ wherein $\boldsymbol{\beta}_{1},\ldots,\boldsymbol{\beta}_{K-1}$
and $m_{j,\boldsymbol{\beta}_{{\displaystyle \cdot}}}({\mathbf t_i})$ for $j\neq k$ are
treated as fixed, such that $\boldsymbol{\eta}_{i}$ is a function only of
$m_{k,\boldsymbol{\beta}_{{\displaystyle \cdot}}}$ in (\ref{smolik}).

With an estimate for $m_{k,\boldsymbol{\beta}_{{\displaystyle \cdot}}}(\cdot)$ at hand, we can
compute the profile log-likelihood
\begin{eqnarray*}\label{prolik}
{\mathcal{L}}_p (\boldsymbol{\beta}_k)   = \sum_{i=1}^n   {\mathcal{L}}
(\boldsymbol{\eta}_{i}(\boldsymbol{\beta}_{k}),y_i), & & 
\\ \nonumber \mbox{where now \, }
\boldsymbol{\eta}_{i}(\boldsymbol{\beta}_{k})  :=(\eta_{1i},\ldots,\eta_{ki}
(\boldsymbol{\beta}_{k}),\ldots,\eta_{Ki}), & & 
\\ \nonumber \mbox{with \, }
\eta_{ki}(\boldsymbol{\beta}_{k})  :={\mathbf x}_i^t \boldsymbol{\beta}_k +
m_{k,\boldsymbol{\beta}_{\cdot}}({\mathbf t_i})
\end{eqnarray*}
and $\eta_{ji}$, $j \neq k$, as before. Note that
$\boldsymbol{\eta}_{i}(\cdot)$ is a function of $\boldsymbol{\beta}_k$ in (\ref{prolik}).

For the estimation procedure we further need the first two derivatives of $
l_i (\boldsymbol{\eta}) = {\mathcal{L}} (\boldsymbol{\eta},y_i) $ with respect to
$\eta_k$, $\eta_k={\mathbf x}_i^t \boldsymbol{\beta}_k + m_{k,\boldsymbol{\beta}_{\cdot}}({\mathbf t_i})$. We have
\begin{equation*}\label{loglik}
l_i (\boldsymbol{\eta}) = \sum_{k=1}^K I_{ \{ y_i=k \}} \eta_{ki}   - \log
\sum_{j=1}^K \exp \left( \eta_{ji} \right) \, , 
\end{equation*}
where $I$ is the indicator function.
 It then follows immediately that
 \begin{eqnarray*}
 l_{ik}'(\boldsymbol{\eta}) &=&
 I_{ \{ y_i=k \}}  - \frac{\exp(\eta_{ki})}{\sum_{j=1}^K \exp(\eta_{ji})} \\
\text{and   } \; l_{ik}''(\boldsymbol{\eta}) &=& - \dfrac{\exp(\eta_{ki})\cdot \sum_{j=1}^K \exp(\eta_{ji})
 - \exp(\eta_{ki})^2}{ ( \sum_{j=1}^K \exp(\eta_{ji}) )^2 }    \, .
 \end{eqnarray*}
To obtain the maximum of the smoothed likelihood ${\mathcal{L}}_s
(m_{k,\boldsymbol{\beta}_{{\displaystyle \cdot}}}({\mathbf t}))$, successively from category 
$1$ to category $K$, we have to solve the first order condition
\begin{equation}\label{meq}
\sum_{i=1}^n (\det \mathbf H)^{-1} \mathcal{K} \bigl( {\mathbf H}^{-1} ({\mathbf t}-{\mathbf t}_i)
\bigr) l_{ik}' \bigl(\boldsymbol{\eta}_{i} ( m_{k,\boldsymbol{\beta}_{{\displaystyle
\cdot}}}({\mathbf t})) \bigr) = 0
\end{equation}
with respect to $m_{k,\boldsymbol{\beta}_{{\displaystyle \cdot}}}({\mathbf t})$. For
$\boldsymbol{\beta}_k$ the equation to solve is
\begin{equation*}\label{beq}
\sum_{i=1}^n  l_{ik}' (\boldsymbol{\eta}_{i}(\boldsymbol{\beta}_{k})) ({\mathbf x}_i+
m_{k,\boldsymbol{\beta}_{{\displaystyle \cdot}}}'({\mathbf t_i})) = \boldsymbol{0} \, ,
\end{equation*}
wherein $m_{k,\boldsymbol{\beta}_{{\displaystyle \cdot}}}'({\mathbf t_i})$ denotes the
gradient of $m_{k,\boldsymbol{\beta}_{{\displaystyle \cdot}}}({\mathbf t_i})$ with
respect to $\boldsymbol{\beta}_k$. By differentiating equation (\ref{meq}) with respect to
$\boldsymbol{\beta}_k$, one obtains
\begin{equation}\label{mpr}
m_{k,\boldsymbol{\beta}_{{\displaystyle \cdot}}}'({\mathbf t})= - \frac{\sum_{i=1}^n (\det
\mathbf H)^{-1} \mathcal{K} \bigl( {\mathbf H}^{-1} ({\mathbf t}-{\mathbf t}_i) \bigr) l_{ik}''
\bigl( \boldsymbol{\eta}_{i}(m_{k,\boldsymbol{\beta}_{{\displaystyle \cdot}}}({\mathbf
t})) \bigr){\mathbf x}_i} {\sum_{i=1}^n (\det \mathbf H)^{-1} \mathcal{K} \bigl( {\mathbf H}^{-1}
({\mathbf t}-{\mathbf t}_i) \bigr) l_{ik}'' \bigl(\boldsymbol{\eta}_{i} (
m_{k,\boldsymbol{\beta}_{{\displaystyle \cdot}}}({\mathbf t}))\bigr)} \, .
\end{equation}
Equations (\ref{meq}) to (\ref{mpr}) can now be used to implement a Newton-Raphson-type
algorithm, involving the following four steps:
 {\it
\begin{itemize}
\item[{1.}] Find appropriate starting values $\boldsymbol{\beta}_k^{(0)}$,
    $m_k^{(0)} (\cdot)$, $k=1, \ldots , K-1$ (e.g.\ by fitting an appropriate
    parametric MNL model) and set $j=0$.
\item[{2.}] For $k=1,2,\ldots,K-1$, compute
   \begin{eqnarray*}
    \boldsymbol{\beta}_k^{(j+1)} &=& \boldsymbol{\beta}_k^{(j)}-{\mathcal{B}}^{-1}
    \sum_{i=1}^{n}    l_{ik}' (\boldsymbol{\eta}_{i}(\boldsymbol{\beta}_{k}^{(j)}))
    ({\mathbf x}_i+m_{k,\boldsymbol{\beta}_{{\displaystyle \cdot}}}'^{(j)}({\mathbf
    t_i}))  \\ \mbox{with } & & {\mathcal{B}}=\sum_{i=1}^{n}    l_{ik}''
    (\boldsymbol{\eta}_{i}(\boldsymbol{\beta}_{k}^{(j)})) ({\mathbf
    x}_i+m_{k,\boldsymbol{\beta}_{{\displaystyle \cdot}}}'^{(j)}({\mathbf
    t_i}))({\mathbf x}_i+m_{k,\boldsymbol{\beta}_{{\displaystyle
    \cdot}}}'^{(j)}({\mathbf t_i}))^t
   \end{eqnarray*}
    and $m_{k,\boldsymbol{\beta}_{{\displaystyle \cdot}}}'^{(j)}({\mathbf t_i})$ as
    in (\ref{mpr}).
\item[{3.}] For $k=1,2,\ldots,K-1$, compute $$
    m_{k,\boldsymbol{\beta}_{{\displaystyle \cdot}}}^{(j+1)}({\mathbf t}) =
    m_{k,\boldsymbol{\beta}_{{\displaystyle \cdot}}}^{(j)}({\mathbf t}) -
    \frac{\sum_{i=1}^n (\det \mathbf H)^{-1} \mathcal{K} \bigl( {\mathbf H}^{-1} ({\mathbf
    t}-{\mathbf t}_i) \bigr) l_{ik}'
    \bigl(\boldsymbol{\eta}_{i}(m_{k,\boldsymbol{\beta}_{{\displaystyle
    \cdot}}}^{(j)}({\mathbf t}))\bigr)}{\sum_{i=1}^n (\det \mathbf H)^{-1} \mathcal{K} \bigl(
    {\mathbf H}^{-1} ({\mathbf t}-{\mathbf t}_i) \bigr) l_{ik}''
    \bigl(\boldsymbol{\eta}_{i}(m_{k,\boldsymbol{\beta}_{{\displaystyle
    \cdot}}}^{(j)}({\mathbf t}))\bigr)}$$ for all points ${\mathbf t}$ at which the
    function $m_{k,\boldsymbol{\beta}_{{\displaystyle \cdot}}}(\cdot)$ is to be
    estimated.
\item[{4.}] Repeat steps {\bf 2.}--{\bf 3.} for $j=1,2,\ldots$ until convergence.
\end{itemize}
}
 It is convenient to estimate the functions $m_{k,\boldsymbol{\beta}_{{\displaystyle
\cdot}}}(\cdot)$ in step 3 at the observation points $\mathbf{t}_i$,
$i=1,\ldots,n$, as this guarantees that independent of the bandwidth choice at least for
one observation $\mathcal{K} \bigl( {\mathbf H}^{-1} ({\mathbf t}-{\mathbf t}_i) \bigr)$ is
nonzero. For different alternatives of implementation see for example Chapter 7 in H{\"a}rdle et al.\ (2004b). 
Category-specific intercepts are not explicitly incorporated into our semiparametric model, since 
the vertical location (not the shape) of the functions $m_k$ accounts for it, and in order to 
be distinguishable from the functions $m_k$ such intercepts would need to vary also over individuals.
In our application the available information on the different political parties
does not vary over individuals.

If the dimension $q$ of variable $\mathbf{T}$ increases, such that 
the curse of dimensionality  becomes an issue (in terms of the estimation performance and with respect to the interpretation
of the nonparametric part), then further assumptions on the structure of the functions $m_k(\mathbf{T})$ may be required.  
The most popular such is additive separability, i.e., modelling 
$$  m_k(\mathbf{T}) = \alpha_k + \sum_{l=1}^q m_{k,l} (T_l)  , \quad k=1,\ldots ,K . $$
Relatively straightforward, though computationally tedious extensions of our approach to such settings could be based on backfitting algorithms or marginal integration techniques. In the profile likelihood context with marginal integration, one could adapt the procedure of H\"ardle et al.\ (2004a). For backfitting, probably the most efficient version is the smooth backfitting for generalized structured models (Roca-Pardinas and Sperlich 2010), which would have to be adapted to multinomial responses. A spline version for additive regression of multicategorical data has been proposed by Tutz and Scholz (2004).

\section{Fitting MNL models to the party affiliation data}\label{results}

\subsection{Results obtained using a fully parametric MNL model}\label{paraMNL}

We initially consider the results of fitting a conventional, fully parametric 
MNL model to the data described in Section \ref{sec-appl1}, incorporating the covariates gender (dummy, 1 if female), region (dummy, 1 if Eastern Germany is the region of origin), log-income and age. 
The main purpose of considering the fully parametric model is that we want to show its deficiencies compared to more flexible models, 
and in particular it will serve as a benchmark when analyzing the results from fitting the semiparametric MNL model (in Section \ref{sec-appl2} below). 
Furthermore, it allows us to easily test the assumption of irrelevant alternatives.
For the parametric MNL model, the estimated coefficients of the log-odds are given in Table \ref{ParMLCoef}.

\begin{table}[htb]
\caption[Table]{Parameter estimates for the parametric MNL model with CU as
the reference category (standard errors in parentheses; ** indicates significance at the 1\%, * at the 5\%, and $\bullet$ at the 10\% level). \\ {\white 0}}\label{ParMLCoef}
{\tabcolsep3pt
\begin{center}
\begin{small}
\begin{tabular}{llllll} & category effect & gender & region & log(income) & age/10 \\
\hline\\[-0.9em]
SPD &{\white -}3.661(0.382)$^{**}$& -0.013(0.051)&-0.088(0.066)&-0.385(0.045)$^{**}$&-0.013(0.002)$^{**}$\\
A90G&{\white -}0.009(0.617)&{\white -}0.307(0.085)$^{**}$&-0.338(0.121)$^\bullet$&{\white -}0.071(0.074)&-0.044(0.003)$^{**}$\\
LP&{\white -}1.811(0.808)$^{*}$&-0.253(0.103)$^{*}$&{\white -}2.641(0.117)$^{**}$&-0.569(0.097)$^{**}$&-0.007(0.003)$^{*}$\\
FDP&-4.762(0.827)$^{**}$&-0.567(0.114)$^{**}$&{\white -}0.235(0.141)$^\bullet$&{\white -}0.496(0.097)$^{**}$&-0.023(0.004)$^{**}$
 \end{tabular}
\end{small}
\end{center}
}
\end{table}

As reference category we used the largest party, i.e., the CU. 
The CU is known to have strong support from older individuals, such that it is not surprising
that the impact of age is found to be significantly negative for all other parties.
Relatively to the reference category,
being from the East substantially raises the likelihood of being affiliated to the LP. 
The green party, A90G, has particularly strong support in the group of female voters, and is not that strongly represented in Eastern Germany.  
Indeed, being a young and female Western German resident is the typical
characterisation of an A90G voter (Walter 2008). 
On average, presence of high income decreases the likelihood of supporting
the LP and the SPD, while it significantly increases that of supporting the FDP. 
The majority of the supporters of the FDP is found to be male.

A possible criticism here is to not have used
a nested (two-level) MNL model, or a multinominal probit model (MNP model), to avoid the assumption of irrelevant alternatives (IAA). 
The obvious argument against the IAA is that voters might first choose between 
either left or conservative parties, and then in a second step decide amongst parties within these groups. 
An argument in favour of the IAA is that for the present German party system it is no longer that obvious to the voters
(as it perhaps was in the past) what exactly a right-left classification implies. 
As a consequence, it is for example by no means clear whether a former voter of the CU switches to the SPD, A90G or the FDP. 
Application of the computationally rather complex MNP model, 
for which it is not clear to us how to implement a semiparametric version,
thus does not seem expedient in the given context. Likewise, we believe it to be difficult to apply a nested MNL model, for which it is essential to correctly predefine  
adequate subsets of parties and their correlation structure. In any case one can test the IAA  
using the Hausman-McFadden test (Hausman and McFadden 1984) or the Small-Hsiao test (Small and Hsiao 1985), at least in the parametric framework. 
We did this for all permutations of the reference category, and repeated the tests including also the squares of log-income and age to see whether the tests reject the 
IAA for more flexible models. In all cases we obtained p-values above 95\%. This confirms 
that IAA is not problematic for the political party affiliation in Germany,
say since 2000, which is an interesting finding in its own right.

\subsection{Results obtained using a semiparametric MNL model}\label{sec-appl2}

We now fit a more flexible model to the party affiliation data, taken from the class of  
semiparametric MNL models that was introduced in Section \ref{sec-model}. 
More specifically, the two dummies corresponding to gender (1 if female) and region (1 if of Eastern German origin) enter the model parametrically, while
the other two covariates, age and log-income, are modelled nonparametrically:
\begin{align*}\label{modelapp}
\nonumber 
\mathbbm{P} & (Y=k \, | \, \text{\it gender, region, log-income, age}) \\ &\qquad = \frac{\exp \left( \beta_{1,k} \cdot \text{\it gender} + \beta_{2,k} \cdot
\text{\it region} + m_k(\text{\it log-income, age})\right)}{\sum_{j=1}^5 \exp \left( \beta_{1,j} \cdot
\text{\it gender} + \beta_{2,j} \cdot \text{\it region} + m_j(\text{\it log-income, age}) \right)  }
\ .
\end{align*}
 
The largest party (CU) was used as reference category (i.e., $\beta_{1,5}=\beta_{2,5}=m_5=0$ with index $j=5$ referring to the CU).
In the nonparametric part a Gaussian kernel was used to construct the local likelihood (\ref{smolik}).
The bandwidth matrix was taken to be diagonal, with smoothing parameters 
chosen from a grid of bandwidths ranging from $0.4$ to $1$ times the standard deviation of age and
log-income, respectively. For the presentation of the results, we will show results obtained using
$0.5$ times the standard deviations of these covariates, 
a value that was chosen based on a visual assessment of the smoothness of the resulting curves. 
Alternatively, one could use a cross-validation approach. 
The estimated parametric effects of the dummies gender and region are listed in Table \ref{SemCoef}. 
They are similar to those obtained for the parametric MNL model, as given in Table \ref{ParMLCoef}.

\begin{table}[!htb]
\caption[Table]{Estimated coefficients, with standard deviations in brackets, for the parametric 
part of the semiparametric MNL model with CU
as the reference category. ** indicates significance at the 1\%, * at the 5\%, and $\bullet$ at the 10\% level. \\ {\white 0}}\label{SemCoef}
\tabcolsep4pt
\begin{center}
\begin{small}
\begin{tabular}{llllll} & gender & region \\
\hline\\[-0.9em]
SPD  & {\white -}0.006(0.047)$^{**}$ & -0.107(0.059)$^\bullet$      \\
A90G & {\white -}0.334(0.081)$^{**}$ & -0.289(0.115)$^{*}$          \\
LP   & -0.217(0.100)$^{*}$           & {\white -}2.630(0.113)$^{**}$ \\
FDP  & -0.561(0.114)$^{**}$          & {\white -}0.251(0.139)$^*$
 \end{tabular}
\end{small}
\end{center}
\end{table}

\begin{figure}[htb]
\includegraphics[width=7.3cm,height=6.5cm]{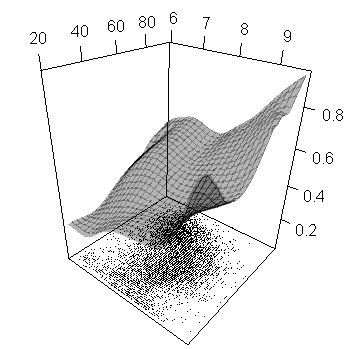}
\includegraphics[width=7.3cm,height=6.5cm]{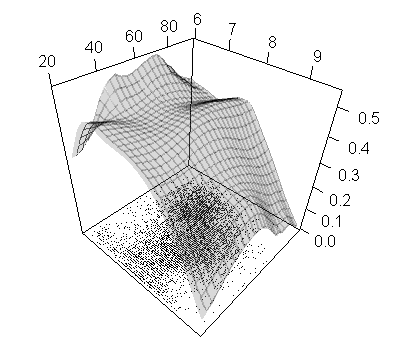}
\includegraphics[width=7.3cm,height=6.5cm]{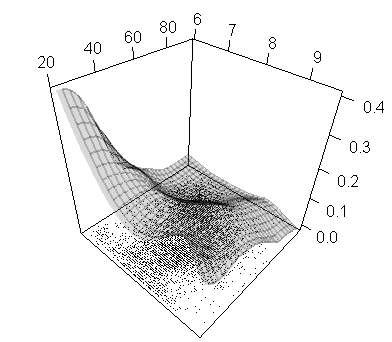}
\includegraphics[width=7.3cm,height=6.5cm]{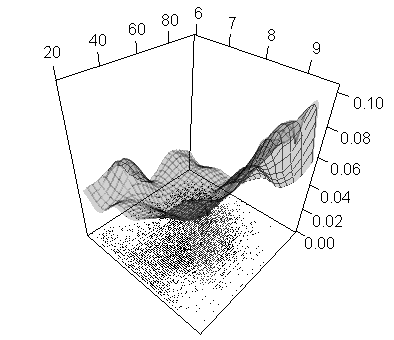}
\caption{Estimated conditional probabilities for women in Western Germany of being a
supporter of the different parties (top left: CU; top right: SPD; bottom left: A90G; bottom right: FDP).
The axis running from 6 to 10 refers to log-income, the one from 20 to 97 refers to age, and the ones restricted to subsets of the interval $[0,1]$ refer to the probability.}
\label{Prob_FW}
\end{figure}

Figure \ref{Prob_FW} displays the
probabilities of supporting any of the parties CU, SPD, A90G and FDP, as a function of age and log-income, for a woman from Western Germany. 
We used the R-packages {\tt rgl} (Adler and Murdoch 2009) and {\tt akima} (based on Akima 1978)
to display the estimated bivariate functions. The use of {\tt akima} generates a smooth surface 
by bivariate interpolation of irregularly spaced input data. 
The black points at the bottom of the plots indicate the observations.
The surfaces have been rotated such that the main features can easily
be recognised. It should be stressed here that for other values of the dummies for gender and region the 
surfaces change (as the probability function is not linear), but only in the sense that some slopes become flatter or steeper; the general patterns remain similar.

\begin{figure}[htb]
\includegraphics[width=7.3cm,height=6.5cm]{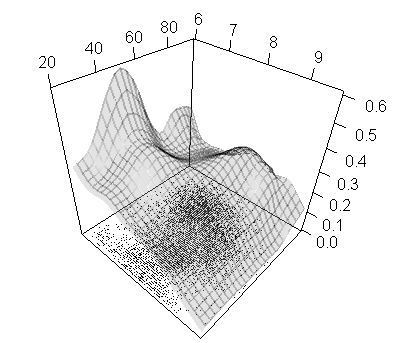}
\includegraphics[width=7.3cm,height=6.5cm]{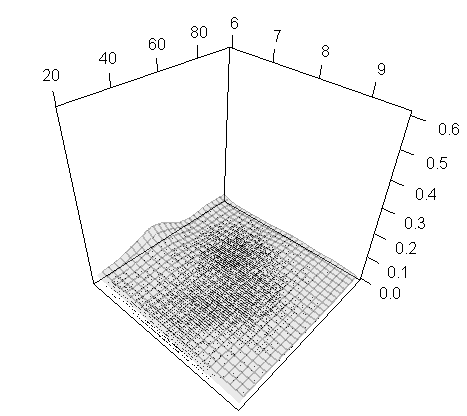}
\caption{Conditional probabilities for men from Eastern (left plot) and from
Western Germany (right plot) of being a supporter of the LP.}
\label{Prob_LP}
\end{figure}

For each party the signs of the slopes change frequently with both increasing age and increasing log-income values. 
It is obvious that both nonlinearities and interactions play an important role for the given voter profiles.
Clearly it would be very difficult to find adequate parametric models that can appropriately reflect 
these important features of the data. In particular, 
it is immediately evident that if the aim is to precisely profile different voter groups, 
then the conventional MNL model, as described in Section \ref{paraMNL}, 
is too simplistic for this type of data.
Even a nonparametric additive decomposition could easily lead to wrong conclusions, since
certain characterisations of voter groups could not be captured by such a model, due to its averaging over all ages to 
derive the influence of the log-income (and vice versa). 
For example, the valley observed for SPD at small log-incomes and young ages would be 
concealed in such an analysis, and likewise, at least to some extent, would be the distinct peak 
observed for LP supporters aged around 50 with very low income (see Figure \ref{Prob_LP}). 
This finding is important, since it implies that most of the standard techniques that are
usually applied to this type of data can  
potentially conceal important data structure.

Considering the surface of probabilities for supporting the CU, we note that the general upward trend in
age is heterogeneous over income: at low incomes the support is essentially monotonically increasing with 
increasing age, but at high incomes ($>3,000$\euro$\,$, i.e., $>8$ on the log scale) this is not the case.
The CU finds stronger levels of support at high income levels than at low income levels.
In contrast, the support of the SPD is strongest in the low income group, over all age groups.
This result corroborates, to some extent, that the SPD is the party of the working class across all ages. For incomes higher than $3,000$\euro$\,$ the likelihood drops away, but
there is a strong support from the middle class with an income of around $3,000$\euro.
The stylized facts for A90G discussed above are found confirmed, but 
with some interesting nuances. The typical voter of the green party 
is generally believed to be either young with low income (mostly students) or relatively well-off (including many civil servants). 
Young individuals with low income indeed constitute a particularly
strong voter block. The expected prominent role of the upper middle class is also found.
However, for ages greater than 50 the latter feature diminishes. 
This is likely due to a strong support from individuals that have an academic background and were involved in 
the seventies' student movements, the eighties' anti-nuclear 
and peace movement in Western Germany, or the peaceful revolution in Eastern Germany. 
The FDP has the smallest voter basis, such that the estimated probability surface in this case is more wiggly and
should not be over-interpreted. However, it can be recognized that 
while age has no clear impact, a high income is most likely to render somebody an FDP supporter. 
Indeed, the impact of log-income on the probability of 
being affiliated to the FDP seems to be exponentially increasing.

Figure \ref{Prob_LP} displays the probabilities of being affiliated to the left party (LP),
for voters from Eastern Germany versus voters from Western Germany, thereby illustrating the effect of the dummy variables. 
We chose the same scales to emphasize the magnitude of the difference between Western and Eastern Germany in this regard:
clearly, the region of origin is the main driving factor that affects whether or not a person is affiliated to the LP.
Focussing on Eastern Germany, we recognize 
strong support in the group of individuals with low income aged between 40 and 60,
whereas a high income (above $6,000$\euro) implies a small probability of an
affiliation with the LP (which is to be expected for a left-wing party). 
We also find a notable support of the LP in the upper middle class for individuals older than 50.
Exactly here one will typically find those who were quite involved in the SED dictatorship, individuals who may have
benefitted from the regime.
 
Overall, the dominant drivers identified by the semiparametric MNL model 
are the same as those already identified by the parametric MNL model (in Section \ref{paraMNL}). 
However, the increase in the flexibility achieved by modelling 
the effects of log-income and age nonparametrically, and by using a bivariate function of those covariates in the predictor, revealed some 
exceptions from the monotonic trends in covariates and led to a number of 
intriguing insights into the interaction between those covariates.

\section{Concluding remarks} \label{sec-concl}

We discussed the utility of a semiparametric MNL model when estimating
political party affiliation. The flexibility of the model leads to comprehensive
insights into the profiles of specific voter groups, insights that other popular modelling approaches 
cannot offer due to inherent simplistic assumptions on the relation between response and explanatory variables.
More specifically, we found that such a model can overcome many potential deficiencies of both 
parametric modelling and nonparametric modelling assuming an additive separability. 
The strong nonlinearities that we found show how complex 
electorates nowadays are structured, and underline the need for more sophisticated 
approaches than parametric MNL modelling, since the nonlinearities would be extremely hard to
capture adequately in a parametric framework.  
Additive modelling, which is designed to allow for nonlinear effects, arguably may 
lead to better interpretability than models involving multidimensional functions of covariates.    
However, in the present application with the given strong interactions between the covariates age and income, 
such a decomposition would still entail a considerable model misspecification and would thus
conceal important structure in the data.
For a detailed discussion on nonparametric additive modelling with 
and without interaction between covariates we refer to Sperlich et al.\ (2002).

Our approach extends the generalized partial linear model (GPLM) framework, as
discussed by M{\"u}ller (2001), to the case of multicategorical responses.
Our methods are in compliance with the GPLM framework, such that the mathematical properties of asymptotic
normality, consistency and efficiency of the estimators remain valid. Our model is directly applicable in other applications 
that are concerned with similar types of data, e.g., in brand choice in marketing studies or regarding the choice of transportation modes.



\end{spacing}

\end{document}